\documentclass[twocolumn,superscriptaddress,prb]{revtex4-2}

\usepackage{array}
\usepackage{graphicx}
\usepackage{dcolumn}
\usepackage{bm}
\usepackage{amsmath}
\usepackage{physics}
\usepackage{bbold}
\usepackage[T1]{fontenc}
\usepackage{dsfont}
\usepackage{tabularx}
\usepackage{comment}
\usepackage{caption}
\usepackage{subcaption}
\usepackage{xcolor}
\usepackage{hyperref}
\usepackage[normalem]{ulem}
\usepackage{natbib}

\renewcommand{\textcolor}[2]{#2}
\renewcommand{\color}[1]{}

\newcommand{\red}[1]{\textcolor{red}{#1}}
\newcommand{\blue}[1]{\textcolor{blue}{#1}}

\newcommand{\jc}[1]{{\color{red}[\textbf{JC: }\textit{{#1}}]}}
\newcommand{\ah}[1]{{\color{blue}[\textbf{AH: }\textit{{#1}}]}}

\begin{document}

\preprint{arxiv}

\title{Majoranas with a twist:\\ Tunable Majorana zero modes in altermagnetic heterostructures}

\author{Andreas Hadjipaschalis}
 \email{andreas.hadjipaschalis@stonybrook.edu}
 \affiliation{Department of Physics and Astronomy, Stony Brook University, Stony Brook, New York 11794, USA}
\author{Sayed Ali Akbar Ghorashi}
\affiliation{Department of Physics and Astronomy, Stony Brook University, Stony Brook, New York 11794, USA}
\author{Jennifer Cano}%
\affiliation{Department of Physics and Astronomy, Stony Brook University, Stony Brook, New York 11794, USA}
\affiliation{Center for Computational Quantum Physics, Flatiron Institute, New York, New York 10010, USA}
\date{\today}

\begin{abstract}


Altermagnetism provides new routes to realize Majorana zero modes with vanishing net magnetization.
We consider a recently proposed heterostructure consisting of a semiconducting wire on top of an altermagnet and with proximity-induced superconductivity.
We demonstrate that rotating the wire serves as a tuning knob to induce the topological phase. 
For $d$-, $g$- and $i$-wave altermagnetic pairing, we derive angle-dependent topological gap-closing conditions.
We derive symmetry constraints on angles where the induced altermagnetism must vanish, which we verify by explicit models.
Our results imply that a bent or curved wire realizes a spatially-dependent topological invariant with Majorana zero modes pinned to positions where the topological invariant changes.
This provides a new experimental set-up whereby a single wire can host both topologically trivial and nontrivial regimes without \textit{in situ} tuning.

\end{abstract}

\maketitle

\section{Introduction}

The pursuit of topological superconductivity (TSC) is driven by the prospect of hosting Majorana zero modes (MZMs) for topological quantum computation \cite{sato2017topological,chiu2016classification,nayak2008nonAbelian,alicea2012new,kitaev2001unpaired}. While the search for an intrinsic topological superconductor is ongoing, significant progress has been made in engineering TSC in 1D and 2D hybrid platforms where magnetism plays a crucial role in breaking TRS \cite{flensberg2021Engineered,lutchyn2018majorana,oreg2010helical,lutchyn2010majorana,sau2010nonAbelian,sau2010genericPlatform,alicea2010majorana,nadjperge2013proposal,qi2010chiralTSC,wang2015chiralTSC,peintka2017topological,shabani2016twodim,schiela2024Shabani,tanaka2024theory}.

One of the most extensively studied approaches to realizing TSC involves 1D heterostructures composed of a semiconducting nanowire with strong spin-orbit coupling (SOC), proximity-induced $s$-wave superconductivity, and an applied magnetic field to break time-reversal symmetry (TRS) \cite{oreg2010helical,lutchyn2010majorana,sau2010nonAbelian}. These three ingredients can cooperate to generate effective $p$-wave topological superconductivity within the wire, leading to the emergence of MZMs localized at its ends, 
as described in Kitaev's seminal paper
\cite{kitaev2001unpaired}. Despite considerable experimental progress probing MZMs in this and related platforms \cite{mourik2012signatures,nadjperge2014observation,deng2016majorana}, two major challenges remain: distinguishing genuine Majorana signatures from those induced by disorder, and mitigating the suppression of the superconducting gap by the TRS-breaking mechanism as a result of the non-zero net magnetization \cite{das2023inSearch,pan2020physical,kayyalha2020absence,liu2017andreev,liu2012zeroBias}. 

Concurrently, efforts to discover novel antiferromagnets with spin-split band structures in the absence of SOC \cite{wu2007,oganesyan2001,noda2016MnO2, okugawa2018, ahn2019pomeranchuk, hayami2019momentumDep,hayami2020bottomUp, naka2019Organic, gonzalez2021spinSplitter, smejkal2022giantMag,yuan2020giantMom,mazin2021FeSb2, Bai2022obsRuo2,karube2022obsRuO2} have led to the emergence of a new magnetic phase, termed altermagnetism \cite{smejkal2022BeyondCon,smejkal2022Emerging}. Altermagnets are typically characterised as collinear, compensated magnets with alternating, anisotropic spin-split bands. These distinctive properties, which combine TRS breaking typical of ferromagnets with the zero net-magnetization typical of antiferromagnets, have been experimentally confirmed, most notably in MnTe \cite{krempasky2024altermagnetic,lee2024brokenKramersMnTe,osumi2024giantSpinSplitMnTe,amin2024nanoscaleAltMnTe} and CrSb \cite{reimers2024altCrSb,zhou2025manipAltCrSb,yang2025threedAltCrSb}, sparking interest across various areas of condensed matter physics, including spintronics \cite{dal2024antiferAndBeyond,chi2024giant,bai2023efficient,zhang2024gateField,gunnink2025acoustic,ezawa2025thirdandfifth,mavani2025switchable,weber2024ultrafast, ghorashi2025dynamicalgeneration}, superconductivity \cite{giil2024SuperAltMem,papaj2023Andreev,sun2023Andreev,maeda2025classSupPairSym,ouassou2023dcJos,zhang2024finiteCoop,zhao2025dSupdAlt,beenakker2023PhaseShift,leraand2025phononmediated,chakraborty2024finiteMom,lu2024Jos,fukaya2025superconducting}, and the anomalous Hall effect \cite{smejkal2020crystalHall,smejkal2022anomalous,reichlova2024obsAnom,gonzalesBetancour2023spontAnomSemi,kluczyk2024Anom,sato2024altAnom,attias2024intrinAnom,sheoran2025twodAHEalt,fang2024quantum}.
\begin{figure}[t!]
\begin{subfigure}{0.61\columnwidth}
    \includegraphics[width=\linewidth]{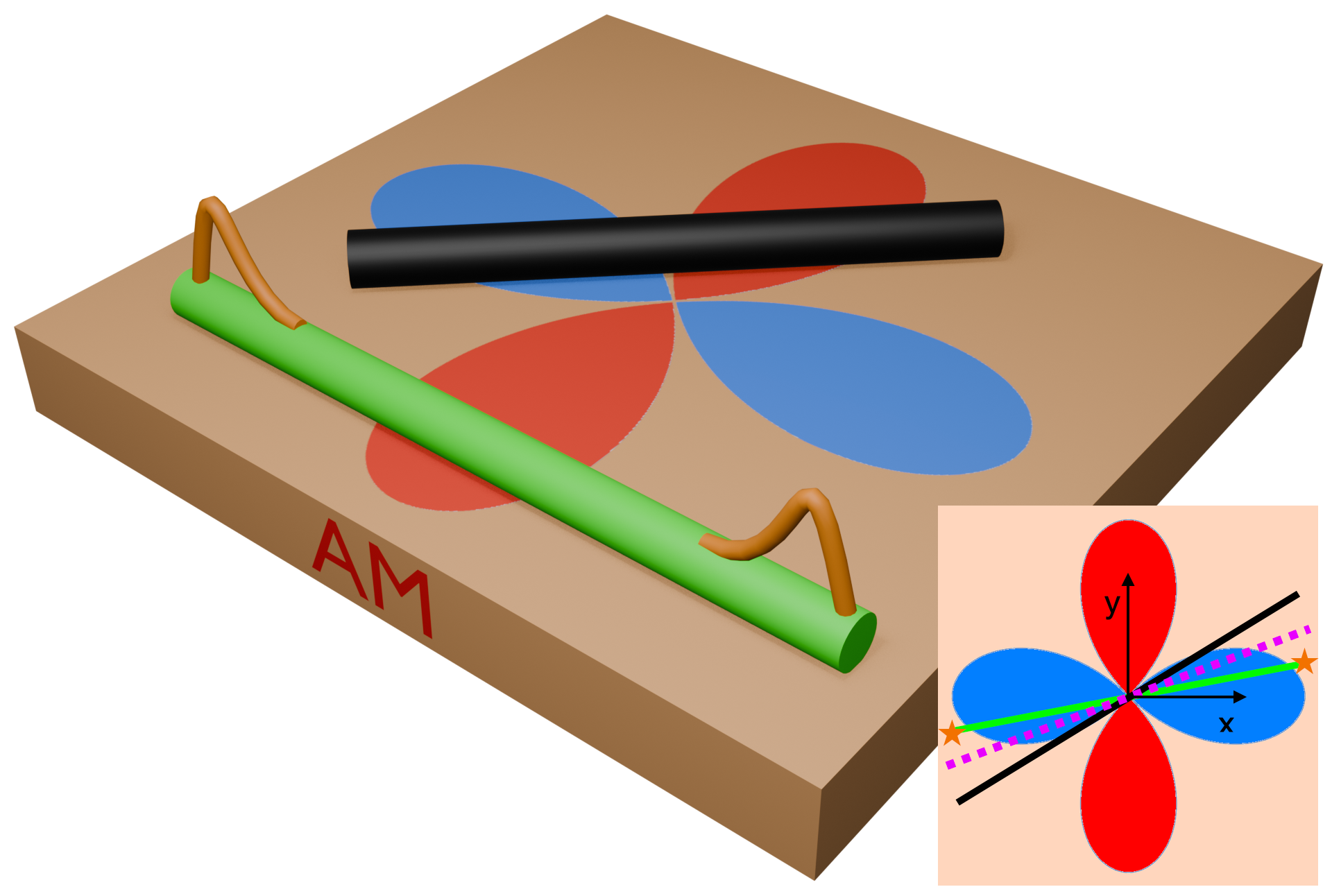}
    \caption{}
    \label{fig:straight}
\end{subfigure}
\begin{subfigure}{0.61\columnwidth}
    \includegraphics[width=\linewidth]{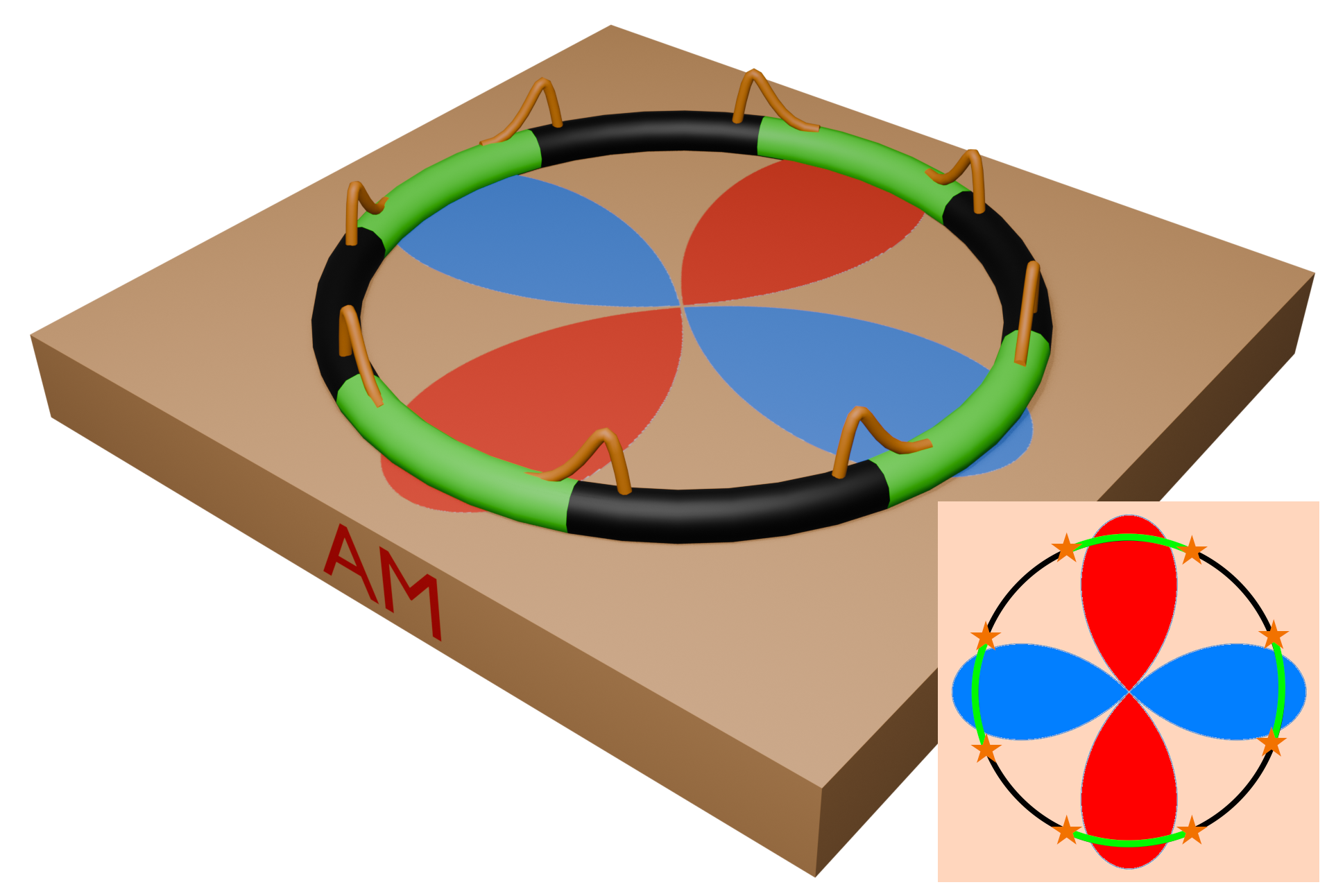}
    \caption{}
    \label{fig:bent}
\end{subfigure}
\caption{Proposed setups consisting of superconducting wires (a) rotated or (b) bent on top of a ($d$-wave) altermagnet, with schematic top-down views in the insets. The red and blue lobes indicate the sign of the underlying altermagnetic order. Green and black segments correspond to topological and trivial regimes respectively, with localized Majoranas at their interfaces (stars in the insets).
In (a), straight wires rotated at different angles relative to the underlying altermagnet reveal that beyond a critical angle (magenta line in inset), a topological phase transition occurs. In (b), a bent wire exhibits a spatially dependent topological order parameter due to its ``continuous rotation''.}
\end{figure}

The unique TRS-breaking mechanism of altermagnets at zero net magnetization also offers a promising route to realizing new intrinsic topological phases \cite{fernandes2024topological,antonenko2025mirror,ma2024altermagneticTop,gonzalez2025spin,li2025altermagnetism,parshukov2024topological,li2024topologicalweyl}. Moreover, altermagnets have been proposed as tools to both enhance and remedy the conventional platforms proposed to realize MZMs \cite{ghorashi2024altermagneticRoute,li2023Majorana,yang2025topologicalalter}. In a recent study involving two of us \cite{ghorashi2024altermagneticRoute}, it was demonstrated that substituting the conventional TRS-breaking mechanism in the 1D heterostructures with proximitized altermagnetism induces the requisite spin splitting in the wire without generating a net magnetization, thereby potentially leading to a larger superconducting gap. Moreover, this approach offers a means to disentangle non-magnetic disorder effects from genuine signatures of Majorana zero modes by the application of an additional, external magnetic field \cite{ghorashi2024altermagneticRoute}.

A key, hitherto unexplored advantage of incorporating altermagnetism into this platform lies in the orientation-dependent nature of altermagnetism. Due to the intrinsically anisotropic spin splitting of altermagnetic order, the proximitized response in the wire is expected to vary with its alignment relative to the underlying altermagnet (Fig. \ref{fig:straight}). Notably, the induced spin splitting should vanish entirely along nodal directions. This effect can be leveraged to tune in and out of the topological phase simply by rotating the wire, distinct from any other TRS-breaking mechanism. It also suggests a potentially powerful diagnostic tool: with all other parameters held constant, true MZMs should appear or disappear at specific critical angles, while disorder-induced signals would remain unaffected. Although this orientation dependence was anticipated \cite{ghorashi2024altermagneticRoute}, it had not been systematically explored.

In this work, we demonstrate that rotating the wire can serve as a novel tuning knob for inducing the topological phase. 
Specifically, we prove that the induced altermagnetism vanishes when the wire is aligned along a nodal line of the altermagnet enforced by mirror symmetry.
We then derive the effective Hamiltonian proposed in \cite{ghorashi2024altermagneticRoute} starting from a microscopic model and extend it to describe a wire sitting at a generic angle on top of a planar altermagnet,
allowing for $d$-wave, $g$-wave, or $i$-wave altermagnetic order. We derive angle-dependent continuum models and, for each order parameter, derive a topological phase diagram in the presence of proximitized superconductivity. The strongest induced altermagnetism, and hence the largest topologically non-trivial region, occurs along the directions of maximum spin splitting.
On the other hand, consistent with our symmetry analysis, the heterostructure is topologically trivial when the wire is aligned along a mirror-symmetry enforced nodal line of the altermagnet, due to vanishing spin splitting.

Our results suggest a promising application: the realization of Majorana zero modes (MZMs) within a wire, rather than solely at its endpoints, by bending the wire into a closed loop (Fig.~\ref{fig:bent}). In this configuration, the wire can be interpreted as undergoing a continuous rotation, leading to the emergence of localized MZMs at specific points where topological phase transitions occur. This mechanism indicates that the platform may host MZMs without the need for additional tuning, potentially streamlining experimental implementation. Moreover, recent experimental advances demonstrating the controlled manipulation of altermagnetic domains \cite{amin2024nanoscaleAltMnTe,zhou2025manipAltCrSb} provide encouraging evidence for the feasibility of the platforms we propose.

\section{Symmetry constraints} \label{SecSymm}
We first prove on symmetry grounds that for a 2D altermagnet with spin pointing out of the plane, the proximitized altermagnetism in the wire vanishes when the wire is aligned along a nodal line of the underlying altermagnet. This effect was observed in \cite{ghorashi2024altermagneticRoute}, but we now prove that it is not specific to any model, but follows generally from the symmetries of the constituent materials.

We restrict ourselves to altermagnets possessing nodal lines that lie in mirror planes perpendicular to the plane of the altermagnet, which guarantees the nodal lines are straight. In the absence of such mirror symmetries, the nodal lines are generically curved, preventing alignment of a straight wire.

Altermagnets are often discussed in the limit of zero SOC within the spin-group formalism, where spin and spatial symmetries are treated independently (see \cite{smejkal2022BeyondCon, jiang2024enumeration, chen2024enumeration,xiao2024spinSpace}, for a detailed discussion of the spin-group formalism and how spin-group symmetries enforce band degeneracies). \blue{However, in the case of out-of-plane spin ordering in 2D, the relevant mirror symmetries protecting the nodal lines
act identically on the spin and orbital degrees of freedom.} Thus, we proceed in the language of standard magnetic point group symmetries. 


\blue{We consider a 1D Rashba wire and keep only the lowest conduction band. The relevant symmetries are time reversal $T$, two vertical mirrors $M_\perp$ and $M_\parallel$ (perpendicular and parallel to the wire), and the two-fold rotation $C_{2z}$ they generate. Their action on a generic term $f(k)\sigma^j$, where $\sigma$ acts on the spin degree of freedom, is given by:}

\blue{\begin{align}
    \mathcal{T}f(k)\sigma^j \mathcal{T}^{-1}&=\begin{cases}
        -f^*(-k)\sigma^j, &\text{if } j=x,y \text{ or } z, \\
        f^*(-k)\sigma^j,  &\text{if } j=\mathbb{1},
    \end{cases} \\
    M_{\perp}f(k)\sigma^jM_{\perp}^{-1}&=\begin{cases}
        -f(-k)\sigma^j, &\text{if } j=y \text{ or } z, \\
        f(-k)\sigma^j,  &\text{otherwise},   
    \end{cases} \\
    M_{||}f(k)\sigma^jM_{||}^{-1}&=\begin{cases}
        -f(k)\sigma^j, &\text{if } j=x \text{ or } z, \\
        f(k)\sigma^j,  &\text{otherwise},   
    \end{cases} \\
    C_{2z}f(k)\sigma^jC_{2z}^{-1}&=\begin{cases}
        -f(-k)\sigma^j, &\text{if } j=x \text{ or } y, \\
       f(-k)\sigma^j,  &\text{otherwise}.   
    \end{cases}
\end{align}}
The most general Hamiltonian respecting these symmetries is given by:
\begin{equation}
    H_{W}=\sum_{k,s,s'}h^{ss'}_{W}(k)c^\dagger_{k,s} c_{k,s'},
\end{equation}
where
\begin{equation}
    h^{ss'}_{W}(k)=t_Wp(k)-\mu_W+\lambda s(k)\sigma^{y,ss'},
    \label{eq:barewiregeneral}
\end{equation}
\blue{and $t_W$ is the hopping strength, $\mu_W$ is the Fermi level of the wire, $\lambda$ is the spin-orbit coupling strength, and $c^{(\dagger)}_{k,s}$ is the annihilation (creation) operator for electrons in the wire with momentum $k$ and spin $s$}. It follows from TRS that $p(k)=p(-k)$ and $s(k)=-s(-k)$. The mirror symmetries forbid terms with \blue{$\sigma^{x}$ or $\sigma^z$}.
Thus, Eq.~(\ref{eq:barewiregeneral}) is consistent with a parabolic dispersion at the conduction band minimum of a semiconductor subject to spin-splitting from Rashba SOC.

Induced altermagnetism with out-of-plane spin order is implemented by a $k$-dependent spin-splitting of the form $a(k)\sigma^z$ with $a(k)=a(-k)$. Such a term breaks TRS -- as expected from coupling to any type of magnet -- but preserves $C_{2z}$, a symmetry of all 2D altermagnets with out-of-plane spin order \cite{smejkal2022BeyondCon}.
Importantly, such a coupling also breaks both mirror symmetries, $M_\perp$ and $M_\parallel$.
If the wire is not aligned along one of the spin-degenerate mirror planes of the altermagnet, these symmetries are broken by the orientation of the wire and hence the induced altermagnetic term is allowed.
However, if the wire is oriented along a mirror plane of the altermagnet, then $M_\perp$ and $M_\parallel$ are symmetries of both the wire and the altermagnet, so they must be preserved in the heterostructure.
Hence, $a(k)$ must vanish and induced altermagnetism is forbidden when the wire is aligned along a nodal line of the altermagnet.

This discussion can be extended to altermagnets with in-plane spin order, which is relevant to the experiments described in Sec.~\ref{SecExp}, but doing so requires the spin-group formalism. Thus, we postpone a more general proof to future work. However, our results in the next section apply to both in-plane and out-of-plane order.

\begin{figure*}[t!]
\centering
\begin{subfigure}{0.32\textwidth}
    \includegraphics[width=\linewidth]{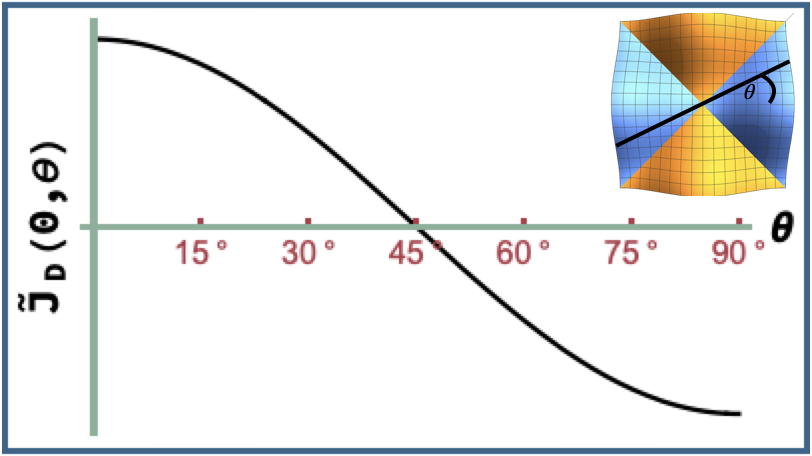}
    \caption{}
    \label{fig:dwave}
\end{subfigure}
\begin{subfigure}{0.32\textwidth}
    \includegraphics[width=\linewidth]{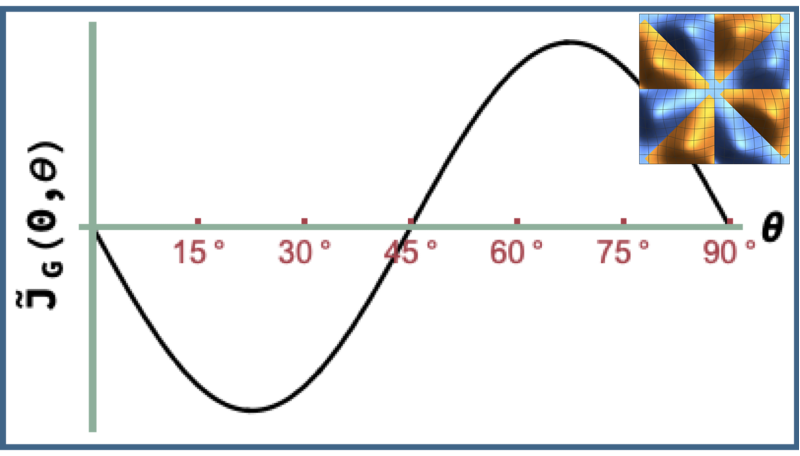}
    \caption{}
    \label{fig:gwave}
\end{subfigure}
\begin{subfigure}{0.32\textwidth}
    \includegraphics[width=\linewidth]{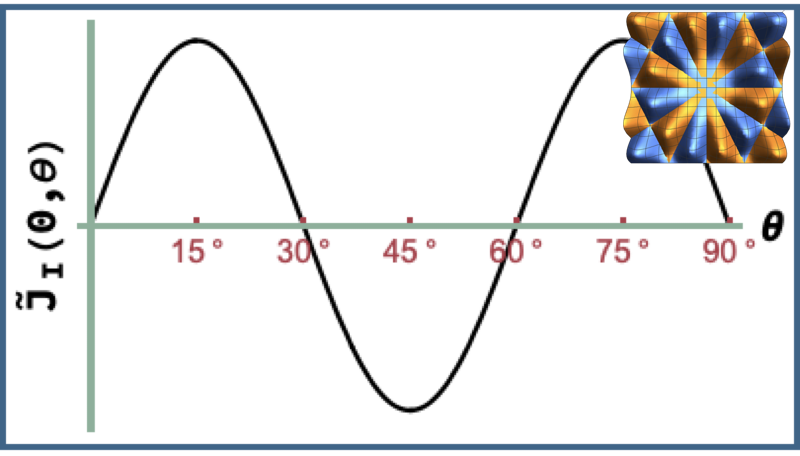}
    \caption{}
    \label{fig:iwave}
\end{subfigure}
\caption{Effective altermagnetic coupling $\Tilde{J}(k,\theta)$ vs angle at $k=0$ for a) $d$-wave, b) $g$-wave and c) $i$-wave altermagnetic order. The insets show the underlying altermagnetic order, with (a) indicating the real space angle of the wire relative to the underlying altermagnetic anisotropy.}
\label{fig:effecCoup}
\end{figure*} 
\section{Microscopic Models}

\begin{table*}[t]
\centering
\begin{tabularx}{0.96\textwidth}{||c| >{\centering\arraybackslash}p{85mm} | >{\centering\arraybackslash}X | c ||}
\hline
 Order & \rule{0pt}{1em}$\Tilde{J}(k,\theta)$ & Topological condition & $t^2_{I,i=D,G,I}$\\
 \hline 
 $d$ & $J\frac{t^2_{I,D}}{\Delta \mu^2}(e^{-\tfrac{1}{2}k^2\epsilon^2}(1-k^2\epsilon^2))\cos(2 \theta)$ & $\frac{t_{I,D}^2}{\Delta\mu^2}J>\sqrt{\Delta^2+\Tilde{\mu}_{W,R}^2}\abs{\sec{2\theta}}$ & $\frac{\epsilon}{8\sqrt{2}\pi^{3/2}}t_I$\\
 \hline
 $g$ & $J\frac{t^2_{I,G}}{\Delta \mu^2}(e^{-\tfrac{1}{2}k^2\epsilon^2}(-3+6k^2\epsilon^2-k^4\epsilon^4)) \sin(4\theta)$ & $\frac{t_{I,G}^2}{\Delta\mu^2}J>\sqrt{\Delta^2+\Tilde{\mu}_{W,R}^2}\abs{\csc{4\theta}}$ & $\frac{1}{\epsilon}\frac{1}{{32\sqrt{2}\pi^{3/2}}}t_I$\\ 
 \hline
 $i$ & $J\frac{t^2_{I,I}}{\Delta \mu^2} (e^{-\tfrac{1}{2}k^2\epsilon^2}(15-45k^2\epsilon^2+15k^4\epsilon^4-k^6\epsilon^6)) \sin(6 \theta)$ & $\frac{t_{I,I}^2}{\Delta\mu^2}J>\sqrt{\Delta^2+\Tilde{\mu}_{W,R}^2}\abs{\csc{6\theta}}$ & $\frac{1}{\epsilon^3}\frac{1}{{16\sqrt{2}\pi^{3/2}}}t_I$\\
 \hline
\end{tabularx}
\caption{\blue{Induced altermagnetism and angle-dependent topological conditions  ($\tilde{\mu}_W=\mu_W-\dfrac{t_I^2}{\Delta\mu}\dfrac{\epsilon^3}{8\sqrt{2}\,\pi^{3/2}}$). }\label{tabResults}}
\end{table*}
We now derive effective Hamiltonians for a spin-orbit coupled wire rotated at a generic angle in proximity to a $d$-wave, $g$-wave, and $i$-wave altermagnet. Our approach is as follows: we weakly couple the wire to the altermagnet and apply a Schrieffer-Wolff (SW) transformation to perturbatively ``integrate out'' the altermagnet. This allows us to determine the angle-dependent gap-closing condition, which in turn enables the construction of the topological phase diagram in the presence of superconductivity. (For a review of the SW transformation, see Appendix \ref{SW}).

To begin, we consider a wire in proximity to a $d$-wave altermagnet and aligned along the $x$-axis. (See the inset to Fig.~\ref{fig:dwave} for the orientation of the altermagnetic order.) Here we sketch the derivation; explicit details can be found in Appendix \ref{lattHamDer}. 
Our calculation provides a microscopic derivation for the Hamiltonian posited in \cite{ghorashi2024altermagneticRoute}. 

We model the wire by considering a special case of \eqref{eq:barewiregeneral} that arises from nearest-neighbor hopping \blue{(with the lattice constant set to 1)}:
\begin{equation}
    h^{ss'}_{W}(k)=t_W \cos{k}-\mu_W+\lambda \sin{k}\,\sigma^{y,ss'}. \label{bareWire}
\end{equation}

We model the $d$-wave altermagnet by the following two-band Hamiltonian:
\begin{align}
    H_{AM}&=\sum_{k,s,s'} h^{ss'}_{AM}(\mathbf{k})d^\dagger_{\mathbf{k},s} d_{\mathbf{k},s'}, \\
    h^{ss'}_{AM}(\mathbf{k})&=\left[ t_{AM}(\cos{k_x}+\cos{k_y})-\mu_{AM} \right] \nonumber
    \\&\qquad+J(\cos{k_x}-\cos{k_y})\sigma^{z,ss'}, \label{dwaveAlt}
\end{align}
where $t_{AM}$ is the hopping strength, $J$ is the altermagnetic spin splitting strength, $\mu_{AM}$ measures the band offset between the wire and altermagnet, \blue{and $d^{(\dagger)}_{\mathbf{k},s}$ corresponds to the electron operator in the altermagnet at momentum $\mathbf{k}=(k_x,k_y)$ and spin $s$}.

For simplicity, we assume the wire and the altermagnet have the same lattice constant and couple them by an on-site $\delta$-function tunneling term (we later relax the lattice constant constraint and consider a more general tunneling function):
\begin{align}
    H_{I}=-t_I\sum_{x} c^\dagger_{x,s}d_{(x,0),s}+\text{h.c.},
    \label{eq:delta}
\end{align}
where $t_I$ is the coupling strength, \blue{and $c^{(\dagger)}_x$ and $d^{(\dagger)}_{x,y}$ are the particle operators in position space in the wire and altermagnet respectively}.

To integrate out the altermagnet, the band alignment must be such that the Fermi level of the wire lies in a band gap of the altermagnet, thereby necessitating that the altermagnet be either insulating or semiconducting. Mathematically, this amounts to $|t_W|,|t_{AM}|, |t_I| \ll  \Delta\mu$, where $ \Delta\mu \equiv |\mu_{AM} - \mu_W| $, facilitating a perturbative expansion.

Upon performing the SW transformation (see Appendix \ref{lattHamDer}), 
we find the following effective Hamiltonian in the wire to leading order in $t_I/\Delta\mu$:
\begin{equation}
	\widetilde{h}^{ss'}_W(k)=\Tilde{t}_W \cos{k}-\Tilde{\mu}_W+\Tilde{\lambda}\sin{k}\sigma^{y,ss'}+\Tilde{J}\cos{k}\sigma^{z,ss'}, \label{wireWithAlt}
\end{equation}
where
\begin{align}
    \Tilde{t}_W&=t_W\left(1-\frac{t_I^2}{\Delta\mu^2}\right)+\frac{t_I^2}{\Delta\mu^2}t_{AM},\\
    \Tilde{\mu}_W&=\mu_W-\frac{t_I^2}{\Delta\mu},\\
    \Tilde{\lambda}&=\lambda\left(1-\frac{t_I^2}{\Delta\mu^2}\right),\\
    \Tilde{J}&=\frac{t_I^2}{\Delta\mu^2}J.
\end{align}
Comparing \eqref{wireWithAlt} to the Hamiltonian for the isolated wire in  \eqref{bareWire}, every parameter has been renormalized and, in addition, there is a new term with $\tilde{J}$ that describes the induced altermagnetism in the wire. As mentioned above, our results provide a microscopic derivation for the Hamiltonian in \cite{ghorashi2024altermagneticRoute} and elucidate how the proximity induced altermagnetic spin-splitting in the wire, parameterized by $\tilde{J}$, is determined from the bare spin-splitting in the altermagnet, parameterized by $J$.


\blue{We next extend the result to a wire rotated by an arbitrary angle $\theta$ on an altermagnet with $d$-, $g$-, or $i$-wave order. Since on-site coupling becomes ill-defined at generic $\theta$, we instead adopt the following continuum models for the wire and altermagnet, avoiding lattice-matching issues:}
\vspace{-0.5em}
\begin{align}
    h^{ss'}_{W,C}(k)&=\frac{k^2}{2m_W}-\mu_{W,C}+\lambda k\,\sigma^{y,ss'}, \\
    h^{ss'}_D(\mathbf{k})&=\frac{k_x^2+k_y^2}{2m_{AM}}-\mu_{AM,C}+J(k_y^2-k_x^2)\sigma^{z,ss'}, \\
    h^{ss'}_G(\mathbf{k})&=\frac{k_x^2+k_y^2}{2m_{AM}}-\mu_{AM,C} \nonumber \\
    &\qquad+J(k_y^2-k_x^2)(k_xk_y)\sigma^{z,ss'}, \\
    h^{ss'}_I(\mathbf{k})&=\frac{k_x^2+k_y^2}{2m_{AM}}-\mu_{AM,C} \nonumber \\
    &\qquad+J(k_xk_y)(3k_y^2-k_x^2)(3k_x^2-k_y^2)\sigma^{z,ss'},
\end{align}
where $m_W$ and $m_{AM}$ are the effective masses of the wire and altermagnet respectively, $\mu_{W,C}$ is the Fermi level in the wire (which differs from $\mu_W$ because it absorbs the constant term in the small-$k$ expansion of the cosine), and $\mu_{AM,C}$ parameterizes the band offset between bands in the wire and the altermagnet via $\Delta \mu = |\mu_{AM,C} - \mu_{W,C}|$. 

The assumptions going into the SW expansion break down due to the arbitrarily high momentum modes in the continuum model. However, these modes are artificial: they arise when the continuum model is used beyond its region of validity. To suppress the unrealistic high-momentum modes, 
we regularize the tunneling between the wire and the altermagnet with a more realistic Gaussian coupling (see Appendix \ref{rotWire} for further discussion): 
\begin{align}
    H_{I,G}=-t_I \int dr dr_{x} dr_{y} e^{-\frac{(r c-r_x)^2+(r s-r_y)^2}{\epsilon^2}} c^\dagger_{r,s} d_{r_x,r_y,s} +\text{h.c.},
\end{align}
where $c=\cos{\theta}$ and $s=\sin{\theta}$, with $\theta$ being the angle of the wire relative to the $x$-axis defined by the altermagnet (see insets to Fig.~\ref{fig:effecCoup}),  and $\epsilon$ sets the width of the Gaussian coupling, which we set to 1.


\begin{figure*}[t!]
\centering
\begin{subfigure}{0.3\textwidth}
    \includegraphics[width=\linewidth]{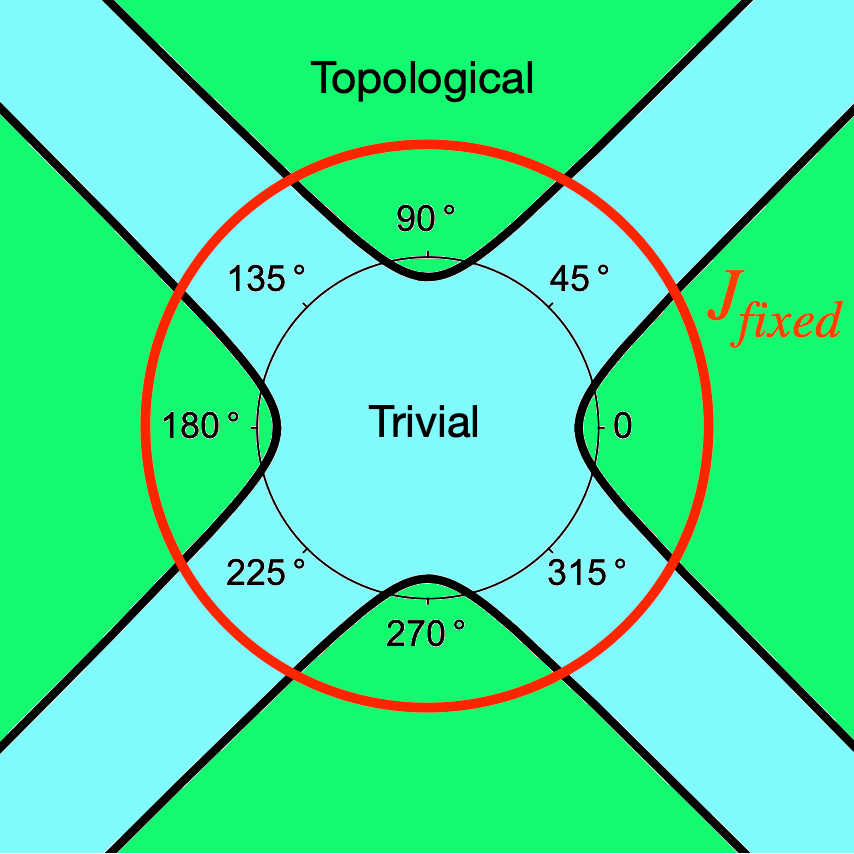}
    \label{fig:figure1}
\end{subfigure}
\begin{subfigure}{0.3\textwidth}
    \includegraphics[width=\linewidth]{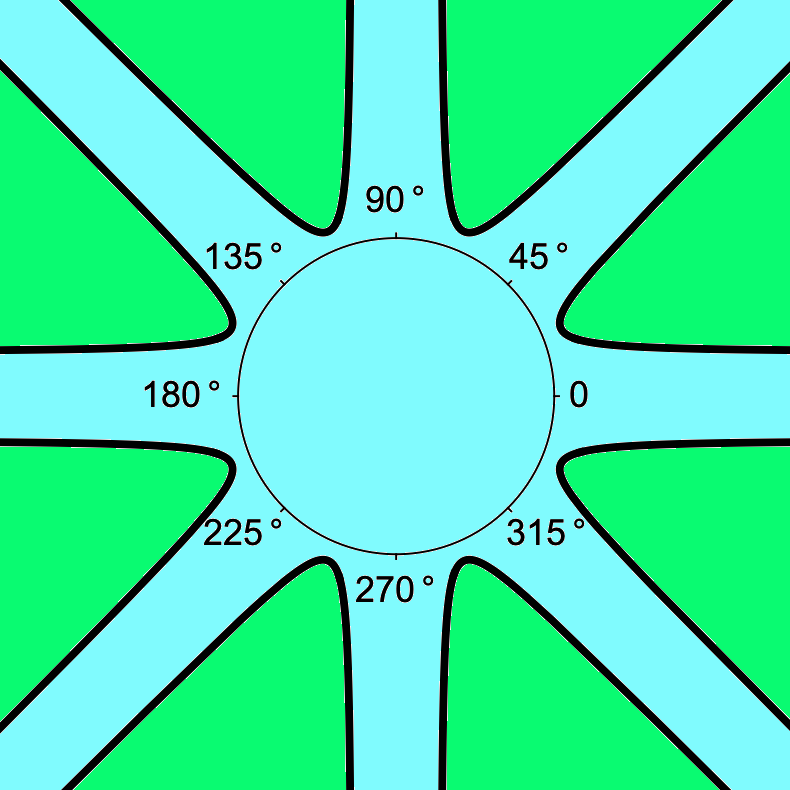}
    \label{fig:figure2}
\end{subfigure}
\begin{subfigure}{0.3\textwidth}
    \includegraphics[width=\linewidth]{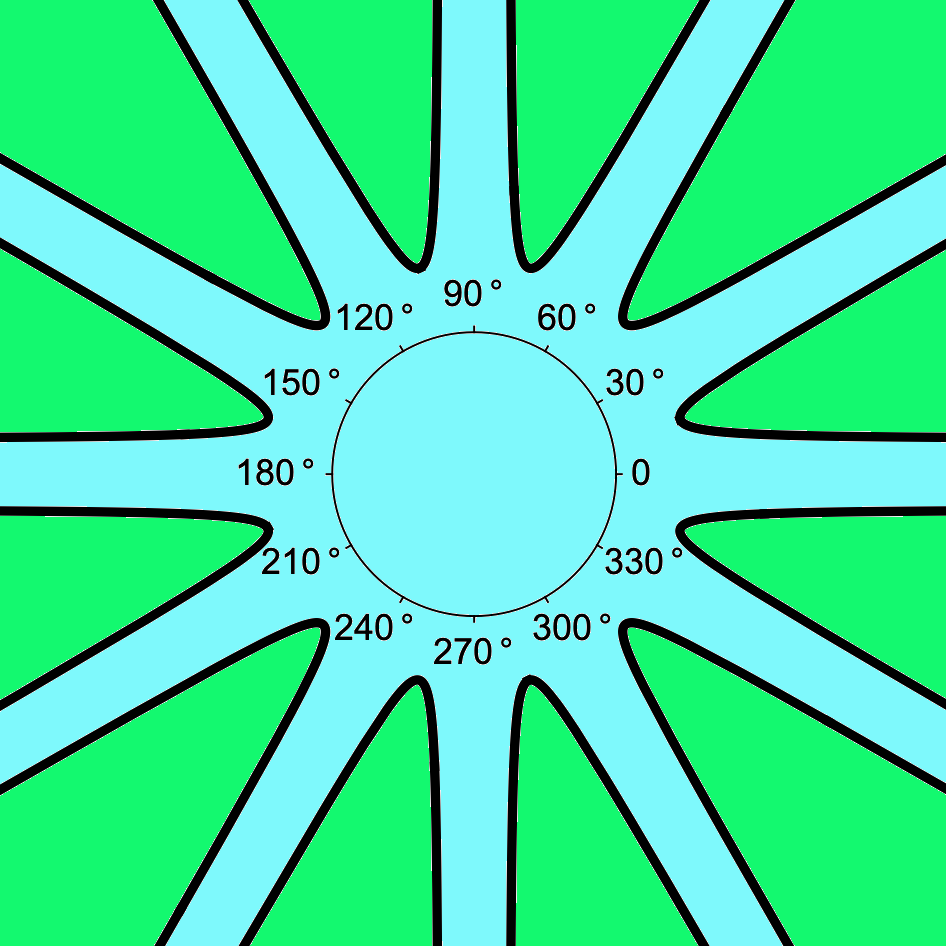}
    \label{fig:figure3}
\end{subfigure}
\caption{Polar plots of topological phase diagrams for a) $d$-wave, b) $g$-wave and c) $i$-wave proximity induced altermagnetism. Radial coordinate corresponds to $J$ from the underlying altermagnet. Plots are made for $\Tilde{\mu}_{W,R}=0.15,$ $\Delta=0.1$ and $\frac{t^2_{I,i}}{\Delta \mu^2}=\frac{1}{25}$ in units of the effective mass with $m_W=1$.} 
\label{figTop}
\end{figure*} 

We can now carry out the SW transformation (see Appendix \ref{rotWire}).
The induced altermagnetism takes the same form as in \eqref{wireWithAlt} but with the coefficient of $\sigma^z$ given in Table \ref{tabResults}. Consistent with our symmetry analysis, we observe that in all cases, the tunneling profile alternates in a manner that mimics the spin splitting of the underlying altermagnet, and, specifically, the induced altermagnetism vanishes when the wire is aligned along the nodal lines, as can be seen in Fig.~\ref{fig:effecCoup}.


Assuming proximity-induced $s$-wave superconductivity (see Appendix \ref{rotWire}), we derive the topological gap closing condition analogous to the condition in conventional setups with an applied magnetic field ($h>\sqrt{\Delta^2+\mu^2}$) \cite{oreg2010helical,lutchyn2010majorana,sau2010nonAbelian,ghorashi2024altermagneticRoute}, but now with an angular dependence (Table \ref{tabResults}).
The result is an angle-dependent topological phase diagram for each altermagnet, shown in Fig.~\ref{figTop}. The largest topologically non-trivial regions occur when the wire is aligned along the directions of greatest spin-splitting, and the topological phases vanish as the wire approaches the nodal lines. Notably, the phase boundaries vary with angle, demonstrating that topological phase transitions are induced by rotating the wire.

\section{Experimental routes to Majorana fermions} \label{SecExp}

We have demonstrated that the relative angle of the wire on top of the altermagnet can be used as a tuning knob to access the topological regime.

We can also apply our results to a curved wire, such as the circular wire shown in Fig.~\ref{fig:bent}. Each small segment of a curved wire can be approximated by a straight wire at a fixed angle, which can be assigned a topological invariant following the appropriate topological phase diagrams in Fig.~\ref{figTop}. The boundaries where the topological invariant changes bind a Majorana fermion, as depicted schematically in Fig.~\ref{fig:bent}. This provides a novel approach to realize different topological regimes within the same wire without requiring external gates or changing the orientation of the wire.

Thus far, our analysis has focused exclusively on altermagnets with a single magnetic domain. However, recent experiments have demonstrated that the altermagnetic order in bulk MnTe can be deliberately engineered to realize altermagnetic domains \cite{amin2024nanoscaleAltMnTe}. In particular, it has been shown that vortex-like nanotextures can be designed and realized within hexagonal microstructures containing multiple magnetic domains, with the Néel vector oriented along the edges of the hexagon. Consequently, placing a nanowire—with an epitaxially grown superconducting shell \cite{liu2020semiconductor,vaitiekėnas2021zeroBias}—across such a microstructure may suffice to generate MZMs without additional tuning. In this configuration, the effective ``rotation'' required for topological phase transitions is provided by the underlying altermagnetic order, rather than by bending the wire itself.

Alternatively, single-domain states have also been realized at the micrometer scale \cite{amin2024nanoscaleAltMnTe}, enabling the implementation of our proposed setup using either bent or straight nanowires. We note, however, that our calculations assume out-of-plane Néel ordering with planar altermagnetism, whereas the experimentally constructed microstructures are in-plane ordered, bulk altermagnets. \blue{The bulk cases are also more subtle, since the induced spin splitting on the surface may be odd in momentum, rather than even, requiring further consideration.} Nevertheless, these experimental developments demonstrating the design and controlled manipulation of altermagnetic order are encouraging. Furthermore, as long as the Néel vector is perpendicular—or possesses a component perpendicular—to the spin-orbit coupling direction in the wire, and as long as the spin splitting has an even component, the qualitative predictions of our analysis are expected to remain valid.

\section{Conclusion}
We have investigated the effects of rotating a one-dimensional wire on the surface of an altermagnet exhibiting $d$-, $g$-, or 
$i$-wave order, as proposed in \cite{ghorashi2024altermagneticRoute}. Our results highlight a novel tuning knob - wire orientation - enabling controlled access to and from the topological regime. 
This gives rise to a new route to realize Majorana fermions \textit{without in situ tuning} by utilizing a bent or curved wire: MZMs are bound to specific, geometry-dependent locations along the wire. Recent experimental demonstrations of controlled manipulation of altermagnetic domains support the feasibility of our approach.

Our findings motivate further studies, particularly first-principles calculations of the heterostructure, to obtain more accurate estimates of the tunneling amplitude between the wire and the altermagnet. In addition, the role of strain in bent wire configurations remains an open question and warrants detailed investigation. Non-collinear altermagnets \cite{Cheong2024altermagnetismNonColl,Cheong2025altermagnetismClas} may also induce spin splitting, potentially giving rise to a richer phase diagram.

Our work underlines the promise of altermagnetic heterostructures in both fundamental research and future quantum technologies, where they may play a pivotal role in the realization and control of exotic quasiparticles.
\section{Acknowledgements} 
We thank Paul Goldbart and Jairo Sinova for enlightening discussions. 
This work is supported by the NSF under Grant No. DMR-1942447,
the Alfred P. Sloan Foundation through a Sloan Research Fellowship and by the Flatiron Institute, a division of the Simons Foundation.

\appendix





\section{Schrieffer-Wolff Transformation} \label{SW}
We provide a brief overview of the Schrieffer–Wolff (SW) transformation as it applies to the present work. Detailed derivations and discussions can be found in several excellent textbooks, e.g., \cite{winkler2003spin,coleman2015introduction}.

The SW transformation derives an effective Hamiltonian in a subspace of the full Hilbert space using quantum mechanical perturbation theory in the Heisenberg (operator) picture.
This is in contrast to perturbation theory in the Schr\"odinger picture where corrections to the unperturbed eigenstates are derived order-by-order in the perturbation.
While the Schr\"odinger picture becomes cumbersome in the presence of degenerate eigenstates, the Heisenberg picture avoids this complication by working at the operator level.

\blue{We consider a Hamiltonian $H = H_0 + \gamma V$, where $H_0$ is the unperturbed Hamiltonian and $\gamma V$ is a weak perturbation ($\gamma\ll 1$). The eigenstates of $H_0$ are assumed to be known and satisfy $H_0 \ket{\psi_n} = E_n \ket{\psi_n}$.}
\blue{In the SW approach, the Hilbert space is divided into two weakly coupled subspaces, $A$ and $B$, separated by an energy gap large compared to the coupling, i.e., $\gamma/(E_m - E_l) \ll 1$ when $\ket{\psi_m}\in A$ and $\ket{\psi_l} \in B$.}
While $H_0$ is diagonal in the unperturbed basis, the perturbation $\gamma V$ mixes states both within and between the subspaces $A$ and $B$.


We seek a unitary transformation $e^S$, where $S$ is an anti-Hermitian matrix, that transforms the unperturbed Hamiltonian $H_0$ into a new Hamiltonian $\Tilde{H}=e^S H e^{-S}$, such that
\begin{equation}
    \bra{\psi_m}\Tilde{H}\ket{\psi_n}= 0 \text{ if }\ket{\psi_m}\in A, \ket{\psi_n}\in B \text{ or vice versa}.
\end{equation}
Determining the unitary transformation at each order amounts to choosing $S=S^{(1)}+S^{(2)}+...$ such that the off-diagonal part of $\gamma V$ vanishes at each order \cite{winkler2003spin}. 
Note that this procedure is agnostic to degeneracy within each subspace.

Up to second order, the SW transformation yields the effective Hamiltonian in the subspace $A$: 
\begin{align}
    \Tilde{H}^{(0)}_{mm'}&=H^0_{mm'}, \label{SW1}\\
    \Tilde{H}^{(1)}_{mm'}&=V_{mm'}, \label{SW2}\\
    \Tilde{H}^{(2)}_{mm'}&=\frac{1}{2}\sum_l V_{ml} V_{lm'}\left[\frac{1}{E_m - E_l} + \frac{1}{E_{m'}-E_l} \right],\label{SW3}
\end{align}
where $V_{mm'}=\bra{\psi_m}\gamma V\ket{\psi_{m'}}$ and similarly for $V_{ml}, V_{lm'}$.


\blue{The practical advantage over Schrödinger-picture perturbation theory is that SW removes the inter-subspace couplings first and only then requires diagonalization within each subspace. In this way it handles degeneracies without having to guess the “correct” unperturbed basis.}

\section{Effective lattice Hamiltonian for a wire with $
\theta=0$ on a $d$-wave altermagnet}\label{lattHamDer}


\blue{We now apply the SW procedure of Appendix~A to the concrete lattice model used in the main text. We take a Rashba wire aligned with the $x$ axis and a $d$-wave altermagnet with the same lattice constant, coupled by on-site tunneling at $y=0$. For clarity we repeat the two Hamiltonians:}
\begin{align}
    h^{ss'}_{W}(k)&=t_W\cos{k}-\mu_W+\lambda\sin{k}\,\sigma^{y,s s'}, \\
    h^{ss'}_{AM}(\mathbf{k})&=\left[ t_{AM}(\cos{k_x}+\cos{k_y})-\mu_{AM} \right] \nonumber
    \\&\qquad+J(\cos{k_x}-\cos{k_y})\sigma^{z,ss'}.
\end{align}
The tunneling coupling is
\begin{align}
    H_I&=-t_I \sum_{s} \sum_{x} c^{\dagger}_{x,s} d_{x,y=0,s} + \text{h.c.}, \label{discCoup}
\end{align}
where, as in the main text, $c_{x,s}$ and $d_{\mathbf{x},s}$ are particle operators at position $x$ ($\mathbf{x}$) and spin $s$ in the wire and altermagnet, respectively.
After Fourier transforming the operators to crystal momentum space:
\begin{align}
    c^{\dagger}_{x,s}&=\frac{1}{\sqrt{N_x}}\sum_{k} e^{-ikx}c^{\dagger}_{k, s}, \\
    d_{x,0,s}&=\frac{1}{\sqrt{N_x N_y}}\sum_{k_x,k_y}e^{ik_x x} d_{k_x,k_y,s},
\end{align}
the tunneling term becomes
\begin{align}
    H_I&= -t_I \frac{1}{N_x\sqrt{N_y}} \sum_{s, x, k, k_x, k_y} e^{i(k_x-k)x} c^\dagger_{k,s} d_{k_x,k_y,s} + \text{h.c.} \nonumber \\
    &=-t_I \frac{1}{\sqrt{N_y}}\sum_{s,k_x,k_y} c^\dagger_{k_x,s} d_{k_x,k_y,s}+\text{h.c},
\end{align}
where $N_x$ and $N_y$ are the total number of lattice sites along the $x$ and $y$ directions respectively ($N_x$ is the same in the wire and altermagnet by assumption) and where we have used $\frac{1}{N_x}\sum_x e^{i(k_x-k)x}=\delta_{k_x,k}$.

As explained in the main text, we take $\Delta \mu \equiv \mu_{AM}-\mu_W$ to be much larger than all other parameters in the theory which allows us to use the SW transformation with the subspace $A$ consisting of states in the wire, $B$ consisting of states in the altermagnet, and $H_I$ being the perturbation i.e. the weak coupling between $A$ and $B$. Since Eqs.~\eqref{SW1}--\eqref{SW3} are derived assuming that the unperturbed basis diagonalizes $H_0$,
we first transform to the spin-$y$ basis in which $H_W$ is diagonal. To this end, we define the spin-y electron annihilation operators $b$ by:
\begin{align}
    c^\dagger_{k,\uparrow}&=\frac{\sqrt{2}}{2}(b^\dagger_{k,\rightarrow}+b^\dagger_{k,\leftarrow}), \\
    c^\dagger_{k,\downarrow}&=\frac{i\sqrt{2}}{2}(-b^\dagger_{k,\rightarrow}+b^\dagger_{k,\leftarrow}),
    \label{eq:spiny}
\end{align}
in terms of which the tunneling Hamiltonian is written as
\begin{align}
    H_I&=-t_I \frac{1}{\sqrt{N_y}} \frac{\sqrt{2}}{2}\sum_{k_x, k_y} (b^\dagger_{k_x,\rightarrow} d_{k_x,k_y,\uparrow} + b^\dagger_{k_x,\leftarrow} d_{k_x,k_y,\uparrow} \nonumber \\
    &\qquad- ib^\dagger_{k_x,\rightarrow} d_{k_x,k_y,\downarrow} + ib^\dagger_{k_x,\leftarrow} d_{k_x,k_y,\downarrow})+\text{h.c.}
\end{align}
The energy eigenvalues of the (unperturbed) wire and altermagnet are:
\\
\begin{align}
    E^{(0)}_{W,k,(\rightarrow,\leftarrow)}&=t_W\cos{k}-\mu_W\pm\lambda\sin{k}, \\
    E^{(0)}_{AM,k_x,k_y,(\uparrow,\downarrow)}&=t_{AM}(\cos{k_x}+\cos{k_y})-\mu_{AM} \nonumber \\
    &\qquad\pm J(\cos{k_x}-\cos{k_y}),
\end{align}
where $+$ and $-$ correspond to spin up and down respectively, remembering that for the wire, spin up and down are taken with respect to the spin-$y$ basis (denoted by right and left arrows, respectively), while for the altermagnet they are in the spin-$z$ basis. 

We now apply \eqref{SW1}--\eqref{SW3} to obtain the effective Hamiltonian for the wire. Towards this end, we calculate the matrix elements $V_{ml}$ where $m=(k,s)$ refers to wire states (subspace $A$) while $l$ refers to altermagnet states (subspace $B$). Note that, in our case, $H^{(1)}_{mm'}=V_{mm'}=0$.
\begin{align}
    V_{ml}&=\bra{k', s_1}H_I\ket{k'_x,k'_y,s_2} \nonumber \\
           &=-t_I\frac{1}{\sqrt{N_y}} \sum_{k_x,k_y} \frac{\sqrt{2}}{2} [\delta_{k',k_x} \delta_{k_x,k'_x} \delta_{k_y,k'_y}(\delta_{s_1,\rightarrow} \delta_{s_2,\uparrow} \nonumber \\&\qquad+\delta_{s_1,\leftarrow}\delta_{s_2,\uparrow} -i\delta_{s_1,\rightarrow}\delta_{s_2,\downarrow}+i\delta_{s_1,\leftarrow}\delta_{s_2,\downarrow})] \label{Velements}.
\end{align}
Therefore (with $m=(k_1,s_1)$ and $m'=(k_3,s_3)$):
\begin{widetext}
\begin{align}
    H^{(2)}_{mm'}=&\delta_{k_1,k_3}\frac{1}{4}\frac{t^2_I}{N_y}\sum_{k_y,s_2}(\delta_{s_1,\rightarrow}\delta_{s_2,\uparrow}+\delta_{s_1,\leftarrow}\delta_{s_2,\uparrow}-i\delta_{s_1,\rightarrow}\delta_{s_2,\downarrow}+i\delta_{s_1,\leftarrow}\delta_{s_2,\downarrow})\nonumber \\
    &\times(\delta_{s_3,\rightarrow}\delta_{s_2,\uparrow}+\delta_{s_3,\leftarrow}\delta_{s_2,\uparrow}+i\delta_{s_3,\rightarrow}\delta_{s_2,\downarrow}-i\delta_{s_3,\leftarrow}\delta_{s_2,\downarrow}) \nonumber \\
    &\times \left[\frac{1}{E^{(0)}_{W(k_1,s_1)}-E^{(0)}_{AM(k_3,k_y,s_2)}}+\frac{1}{E^{(0)}_{W(k_1,s_3)}-E^{(0)}_{AM(k_3,k_y,s_2)}}\right] \nonumber \\
    =&\delta_{k_1,k_3}\frac{1}{4}\frac{t^2_I}{N_y} \sum_{k_y,s_2}(\delta_{s_1,\rightarrow}\delta_{s_2\uparrow}\delta_{s_3,\rightarrow}+\delta_{s_1,\rightarrow}\delta_{s_2\uparrow}\delta_{s_3,\leftarrow}+\delta_{s_1,\leftarrow}\delta_{s_2\uparrow}\delta_{s_3,\rightarrow}+\delta_{s_1,\leftarrow}\delta_{s_2\uparrow}\delta_{s_3,\leftarrow} \nonumber \\
    &+\delta_{s_1,\rightarrow}\delta_{s_2\downarrow}\delta_{s_3,\rightarrow}-\delta_{s_1,\rightarrow}\delta_{s_2\downarrow}\delta_{s_3,\leftarrow}-\delta_{s_1,\leftarrow}\delta_{s_2\downarrow}\delta_{s_3,\rightarrow}+\delta_{s_1,\leftarrow}\delta_{s_2\downarrow}\delta_{s_3,\leftarrow})\nonumber \\&\times\left[\frac{1}{E^{(0)}_{W(k_1,s_1)}-E^{(0)}_{AM(k_3,k_y,s_2)}}+\frac{1}{E^{(0)}_{W(k_1,s_3)}-E^{(0)}_{AM(k_3,k_y,s_2)}}\right]. \label{integrals}
\end{align}
\end{widetext}
As a result of $\delta_{k_1,k_3}$, which enforces translation invariance in the wire, we define $k \equiv k_1=k_3$ and omit the Kronecker delta in the remaining calculations.
Defining
\begin{equation}
    A(s_1,s_2)=\frac{1}{E^{(0)}_{W(k,s_1)}-E^{(0)}_{AM(k,k_y,s_2)}},
\end{equation}
and further simplifying yields:
\begin{widetext}
\begin{align}
    H^{(2)}_{mm'}=\frac{1}{4}&\frac{t^2_I}{N_y} \sum_{k_y}[\delta_{s_1,\rightarrow}\delta_{s_3,\rightarrow}(A(\rightarrow,\uparrow)+A(\rightarrow,\uparrow)+A(\rightarrow,\downarrow)+A(\rightarrow,\downarrow))+
    \delta_{s_1,\rightarrow}\delta_{s_3,\leftarrow}(A(\rightarrow,\uparrow)+A(\leftarrow,\uparrow)-A(\rightarrow,\downarrow)-A(\leftarrow,\downarrow)) \nonumber \\
    &+\delta_{s_1,\leftarrow}\delta_{s_3,\rightarrow}(A(\leftarrow,\uparrow)+A(\rightarrow,\uparrow)-A(\leftarrow,\downarrow)-A(\rightarrow,\downarrow)) +
    \delta_{s_1,\leftarrow}\delta_{s_3,\leftarrow}(A(\leftarrow,\uparrow)+A(\leftarrow,\uparrow)+A(\leftarrow,\downarrow)+A(\leftarrow,\downarrow))].
\end{align}
\end{widetext}
We now transform to the spin-$z$ basis of the wire by noting that $\delta_{s_1,(\rightarrow,\leftarrow)}\delta_{s_3,{(\rightarrow,\leftarrow)}}$ corresponds to $b^\dagger_{k,(\rightarrow,\leftarrow)} b_{k,(\rightarrow,\leftarrow)}$, and the inverse of \eqref{eq:spiny}) yields $b^\dagger_{k,\rightarrow}=\frac{1}{\sqrt{2}}(c^\dagger_{k,\uparrow}+ic^\dagger_{k,\downarrow})$, $b^\dagger_{k,\leftarrow}=\frac{1}{\sqrt{2}}(c^\dagger_{k,\uparrow}-ic^\dagger_{k,\downarrow})$. However, $s_1$ in $A(s_1,s_2)$ still refers to the spin in the spin-$y$ basis since this is the eigenbasis of the unperturbed wire. Grouping in terms of Pauli matrices, we find:
\begin{align}
    H^{(2)}_{mm'}=\frac{1}{4}&\frac{t^2_I}{N_y}\sum_{k_y}[\mathbb{1}(A(\rightarrow,\uparrow)+A(\rightarrow,\downarrow)+A(\leftarrow,\uparrow)\nonumber \\&+A(\leftarrow,\downarrow))+ \nonumber \\ 
    &\sigma^z(A(\rightarrow,\uparrow)-A(\rightarrow,\downarrow)+A(\leftarrow,\uparrow)-A(\leftarrow,\downarrow))+ \nonumber \\
    &\sigma^y(A(\rightarrow,\uparrow)+A(\rightarrow,\downarrow)-A(\leftarrow,\uparrow)-A(\leftarrow,\downarrow))]. \label{discrFin}
\end{align}
We now rewrite this term in the effective Hamiltonian explicitly in terms of the parameters in the unperturbed Hamiltonian by expanding the coefficients $A$ to leading order:
\begin{align}
A(\rightarrow,\uparrow)+A(\rightarrow,\downarrow)+A(\leftarrow,\uparrow)&+A(\leftarrow,\downarrow)\cong \nonumber \\ -\frac{t_W}{\Delta\mu^2}4\cos{k}+\frac{t_{AM}}{\Delta\mu^2}4&(\cos{k}+\cos{k_y})-\frac{4}{\Delta\mu}, \label{tCor} \\
    A(\rightarrow,\uparrow)-A(\rightarrow,\downarrow)+A(\leftarrow,\uparrow)&-A(\leftarrow,\downarrow)\cong \nonumber \\ &\frac{4}{\Delta\mu^2}J(\cos{k}-\cos{k_y}), \label{JCor}\\
    A(\rightarrow,\uparrow)+A(\rightarrow,\downarrow)-A(\leftarrow,\uparrow)&-A(\leftarrow,\downarrow)\cong \nonumber \\&-\frac{4}{\Delta\mu^2}\lambda\sin{k}. \label{lambdaCor}
\end{align}
The sum over $k_y$ in \eqref{discrFin} eliminates the $\cos{k_y}$ terms, so that finally we arrive at the effective Hamiltonian:
\begin{equation}
    \widetilde{h}^{ss'}_W(k)=\Tilde{t}_W \cos{k}-\Tilde{\mu}_W+\Tilde{\lambda}\sin{k}\sigma^{y,ss'}+\Tilde{J}\cos{k}\sigma^{z,ss'},
\end{equation}
where
\begin{align}
    \Tilde{t}_W&=t_W\left(1-\frac{t_I^2}{\Delta\mu^2}\right)+\frac{t_I^2}{\Delta\mu^2}t_{AM},\\
    \Tilde{\mu}_W&=\mu_W -\frac{t_I^2}{\Delta\mu},\\
    \Tilde{\lambda}&=\lambda\left(1-\frac{t_I^2}{\Delta\mu^2}\right),\\
    \Tilde{J}&=\frac{t_I^2}{\Delta\mu^2}J.
\end{align}
\section{Continuum models for rotated wires} \label{rotWire}
One natural course of action would be to repeat the calculation on the lattice for $\theta\neq0$. However, rigidly enforcing commensurability may lead to dramatically different length scales between the wire and altermagnet, in particular at small angles, which is unrealistic. Moreover, the mismatch between the lattice constants introduces complications due to the appearance of non-zero reciprocal lattice vectors in sums of the form:
\begin{equation}
    \sum_r e^{i\Tilde{k}ra}=\frac{2\pi}{a}\sum_q \delta(\Tilde{k}+q), \label{recip}
\end{equation}
where $a$ is the lattice constant of the wire, $q$ represents a reciprocal lattice vector, and $\Tilde{k}$ is a linear combination of momenta in the wire and altermagnet, depending on the angle. Each angle must be considered separately, with different reciprocal lattice vectors contributing at each angle, making it difficult to generalise the calculation for all angles.

We handle these complications by using a low energy continuum model.
\blue{In such a model, as long as we are in the parameter regime close to where the gap closes at $k=0$, and not at $k=\pi/a$, our results will be valid.}

This introduces a new complication: integrating over arbitrarily high momentum modes in the SW procedure violates the assumption of a large energy gap between the wire and altermagnet. However, this complication is an artifact of pushing the continuum model beyond its regime of validity and can be remedied by cutting off or suppressing the high momentum modes. A natural way to do this is to change from an on-site coupling to a Gaussian coupling:
\begin{align}
    H_I=-t_I \sum_s \int dr dr_{x} dr_{y} e^{-\frac{(r c-r_x)^2+(r s-r_y)^2}{\epsilon^2}} c^\dagger_{r,s} d_{r_x,r_y,s} +\text{h.c.},
\end{align}
where $c=\cos{\theta}$ and $s=\sin{\theta}$. Not only does this coupling term serve to suppress high momentum modes, it is also a more realistic coupling where an electron on an atom in the wire can tunnel into a state on any nearby atom in the altermagnet.

Fourier transforming this term yields:
\begin{align}
H_I=-t_I \frac{\epsilon^2}{4 \pi} \int_{-\infty}^\infty dk_x dk_y e^{-\frac{\epsilon^2}{4}(k_x^2+k_y^2)}c^\dagger_{k_xc+k_y s}d_{k_x,k_y}+\text{h.c.},
\end{align}
where $c=\cos{\theta}$ and $s=\sin{\theta}$.

\blue{The continuum Hamiltonians for the wire, $d$-wave, $g$-wave and $i$-wave altermagnets are repeated here for clarity}:
\begin{align}
    h^{ss'}_{W,C}(k)&=\frac{k^2}{2m_W}-\mu_{W,C}+\lambda k\sigma^{y,ss'}, \\
    h^{ss'}_D(\mathbf{k})&=\frac{k_x^2+k_y^2}{2m_{AM}}-\mu_{AM,C}+J(k_y^2-k_x^2)\sigma^{z,ss'}, \\
    h^{ss'}_G(\mathbf{k})&=\frac{k_x^2+k_y^2}{2m_{AM}}-\mu_{AM,C} \nonumber \\
    &\qquad+J(k_y^2-k_x^2)(k_xk_y)\sigma^{z,ss'}, \\
    h^{ss'}_I(\mathbf{k})&=\frac{k_x^2+k_y^2}{2m_{AM}}-\mu_{AM,C} \nonumber \\
    &\qquad+J(k_xk_y)(3k_y^2-k_x^2)(3k_x^2-k_y^2)\sigma^{z,ss'}.
\end{align}
We proceed to apply the SW transformation in exactly the same way as in Appendix~\ref{lattHamDer}, with sums replaced by integrals \blue{and discrete Kronecker deltas $\delta_{k',k_x}$ replaced by continuous $\delta$-functions} $\delta(k'-k_x\cos\theta-k_y\sin\theta)$ in \eqref{Velements}. 
Note that the integrals in \eqref{integrals} should only be performed after expanding to linear order as in \eqref{tCor}-\eqref{lambdaCor}. Otherwise, the unphysical singularities discussed earlier will appear.

\blue{For each order of altermagnetism, we find the following proximitized responses:}
\begin{align}
    \Tilde{J}_D(k,\theta)&=J\frac{t^2_{I,D}}{\Delta\mu^2}e^{-\frac{1}{2}k^2\epsilon^2}(1-k^2 \epsilon^2)\cos{2\theta}, \label{A} \\
    \Tilde{J}_G(k,\theta)&=J\frac{t_{I,G}^2}{\Delta\mu^2}e^{-\frac{1}{2}k^2\epsilon^2} \nonumber \\&\times(-3 + 6 k^2 \epsilon^2 - k^4 \epsilon^4)\sin{4 \theta}, \label{B} \\
    \Tilde{J}_I(k,\theta)&=J\frac{t^2_{I,I}}{\Delta \mu^2}e^{-\frac{1}{2}k^2\epsilon^2}\nonumber \\ &\times(15-45k^2\epsilon^2+15k^4\epsilon^4-k^6\epsilon^6)\sin{6 \theta}, \label{C}
\end{align}
where we have absorbed constants into $t^2_{I,D}=\frac{\epsilon}{8\sqrt{2}\pi^{3/2}}t_I$, $t^2_{I,G}=\frac{1}{\epsilon}\frac{1}{{32\sqrt{2}\pi^{3/2}}}t_I$ and $t^2_{I,I}=\frac{1}{\epsilon^3}\frac{1}{{16\sqrt{2}\pi^{3/2}}}t_I$. 

The effective mass $m_W$, spin-orbit coupling $\lambda$ and chemical potential $\mu_W$ in the wire get renormalized in the same way, regardless of the underlying altermagnetic order:
\begin{align}
    \Tilde{m}_W&=m_W\left(1+\frac{t_I^2}{\Delta \mu^2}\frac{\epsilon^3}{8\sqrt{2}\pi^{3/2}}-\frac{m_W}{m_{AM}}\frac{t_I^2}{\Delta \mu^2} \frac{\epsilon^3}{16\sqrt{2}\pi^{3/2}}\right),\\
    \Tilde{\mu}_W&=\mu_W-\frac{t^2_I}{\Delta\mu}\frac{\epsilon^3}{8\sqrt{2} \pi^{3/2}},\\
    \Tilde{\lambda}&=\lambda\left(1-\frac{t^2_I}{\Delta\mu^2}\frac{\epsilon^3}{8\sqrt{2} \pi^{3/2}}\right),
     \\ \nonumber
\end{align}
where we have expanded the exponential and collected the coefficients of $k^2,k$ and the constant term respectively.

Superconductivity can then be included in the wire using the Bogoliubov-de Gennes formalism (suppressing indices and particle operators):
\begin{align}
    H_{W,S}=\left(\frac{k^2}{2\Tilde{m}_R}-\Tilde{\mu}_{W,R}+\Tilde{\lambda}_Rk\sigma^{y}+\Tilde{J}_i(k,\theta)\sigma^{z}\right)\tau^z
    +\Delta\sigma^{y}\tau^y, \label{wireSC}
\end{align}
where $\tau$ correspond to the Nambu space. The subscript $R$ indicates that the couplings are renormalized in addition to the altermagnet by the superconductor. The proximitized altermagnetic term $\tilde{J}$ is also renormalized, but at a higher order (resulting from an electron in the wire tunneling into the altermagnet and back, and then into the superconductor and back), so we have omitted the subscript $R$ on that term.
It is then straightforward to diagonalize the above Hamiltonian and determine the gap closing conditions at $k=0$, as presented in the main text.

\bibliography{main}
\end{document}